\documentclass[twocolumn]{aastex6}

\begin{document}

\title{The dependence of cluster galaxy properties on the central entropy of their host cluster}

\author{Jae-Woo Kim\altaffilmark{1}, Jongwan Ko\altaffilmark{1,2}, Ho Seong Hwang\altaffilmark{3}, 
Alastair C. Edge\altaffilmark{4}, Joon Hyeop Lee\altaffilmark{1,2}, Jong Chul Lee\altaffilmark{1}, 
and Hyunjin Jeong\altaffilmark{1}}
\email{kjw0704@kasi.re.kr}

\altaffiltext{1}{Korea Astronomy and Space Science Institute, 776 Daedeokdae-ro, Yuseong-gu, Daejeon 34055, Republic of Korea}
\altaffiltext{2}{University of Science and Technology, Daejeon 34113, Republic of Korea}
\altaffiltext{3}{School of Physics, Korea Institute for Advanced Study, 85 Hoegiro, Dongdaemun-gu, Seoul 02455, Republic of Korea}
\altaffiltext{4}{Centre for Extragalactic Astronomy, Durham University, South Road, Durham DH1 3LE, UK}



\begin{abstract}

We present a study of the connection between brightest cluster galaxies (BCGs) and their host galaxy clusters.
Using galaxy clusters at $0.1<z<0.3$ from 
the Hectospec Cluster Survey (HeCS) with X-ray information from the Archive of 
{\it Chandra} Cluster Entropy Profile Tables (ACCEPT), we confirm that BCGs
in low central entropy clusters are well aligned with the X-ray center. 
Additionally, the magnitude difference between BCG and the 2nd 
brightest one also correlates with the 
central entropy of the intracluster medium. 
From the 
red-sequence (RS) galaxies, we cannot find significant dependence of RS color scatter 
and stellar population on the central entropy of the intracluster
medium of their host cluster. However, BCGs in low entropy clusters 
are systematically less massive than those in high entropy clusters, 
although this is dependent on the method used to derive the stellar 
mass of BCGs. 
In contrast, the stellar velocity dispersion of BCGs shows no dependence  
on BCG activity and cluster central entropy. This implies that the  
potential of the BCG is established earlier and 
the activity leading to optical emission lines is dictated by 
the properties of the intracluster medium in the cluster core.

\end{abstract}

\keywords{galaxies: clusters: general --- galaxies: elliptical and lenticular, cD --- 
galaxies: formation --- galaxies: evolution --- galaxies: kinematics and dynamics}

\section{INTRODUCTION} \label{intro}

Under the $\Lambda$ cold dark matter ($\Lambda$CDM) paradigm, the dark matter halo evolves hierarchically from 
small density fluctuations to large cluster-like structures. A galaxy cluster represents the most massive dark matter halo or a density peak in the universe. 
In addition, since it is also believed that galaxies form and evolve at the center of dark matter halos \citep{whi78}, the formation and 
evolution of galaxies are significantly affected by the property of host dark matter halos \citep[e.g.,][]{bau06}.

In this context, the correlations between properties of brightest cluster galaxies (BCGs), usually located at the center of galaxy clusters, and their 
host galaxy clusters allow us to understand the environmental effect of the host clusters on the formation and evolution of the 
central galaxies. 
However, many results have reported that the position of BCG and the center of 
X-ray emission is not always coincident. Also, the amount of offset is well correlated with the property of galaxy clusters. BCGs in galaxy clusters 
with low central entropy values or cooling flow are well aligned with the X-ray centers, but those with high central entropy or distorted X-ray morphology show the larger 
offset between them \citep{san09,hof12,gro14,has14}. 

Another interesting aspect of BCGs is their star formation and nuclear
activity. BCGs are regarded as the most massive galaxies in the universe and
hence host the most massive Black Holes and largest stellar populations. 
However, the fraction of BCGs show emission lines in their spectra or blue optical colors, which are different from the widely adopted properties of 
red, passive early-type galaxies at low redshift. 
The BCGs with emission lines (active BCGs, hereafter) usually show a small separation from the X-ray center \citep{cra99,san09}, and reside in 
galaxy clusters with low central entropy or short cooling time, that are considered as cool-core clusters \citep{cav08,raf08,wan10,pip11,hof12,fog15}. 
They also show color excesses at UV and mid-IR regimes \citep{don10,hof12,gre16}. Moreover, it is known that active BCGs show dust \citep{edg99,raw12}, warm 
molecular hydrogen \citep{edg02,ega06}, CO emission \citep{edg01,sal03}, and atomic cooling lines \citep{edg10,mit12}.
All this evidence supports the star formation activity or the existence of active galactic nuclei (AGN) in active BCGs, and these are related to 
the intracluster medium of host clusters.

It is expected that the comparison between BCGs and other cluster galaxies provides an opportunity to understand the evolutionary stage of galaxy clusters or 
the evolution history of BCGs, since BCGs are dominant galaxies in galaxy clusters in terms of brightness and mass \citep{ost75,loh06}. 
For example, \citet{gre16} find that active BCGs tend to show larger magnitude 
difference from the 2nd brightest one than passive BCGs without emission lines, which is consistent with \citet{lau14}. 
However, studies are still not enough to draw a firm conclusion regarding the correlations between BCGs and other cluster galaxies due to the 
lack of deep spectroscopy confirming the enough number of member galaxies. 

As already mentioned, since galaxies evolve in their host dark matter halos, the stellar mass (or luminosity) of central galaxies is tightly connected to 
the halo mass \citep{zhe07,beh10,mos10,wak11,kim15}. In addition, the relation between them is related to the efficiency of conversion from baryons to stars, and important to 
constrain galaxy formation and evolution models. The galaxy clusters are also good laboratories to directly measure the stellar mass-halo mass ratio 
\citep{lin04,gon07,han09,kra14,hwa16}. 

In this work, we use galaxy clusters at $0.1<z<0.3$ from the Hectospec Cluster Survey \citep[HeCS,][]{rin13} with X-ray information from the Archive of {\it Chandra} 
Cluster Entropy Profile Tables \citep[ACCEPT,][]{cav09}. This is one of the best datasets to study the interrelation between member galaxies and their host clusters, 
since we can use a large number of spectroscopically confirmed members with broad-band photometry and uniformly analyzed X-ray information. Using this, we investigate 
the dependence of the property of cluster galaxies on that of host clusters, and address the relation between masses of central galaxies and 
galaxy clusters.

In \S~\ref{data}, we introduce datasets about cluster sample, photometry/spectroscopy and properties of galaxies and clusters. Then 
main results and discussion are presented in \S~\ref{resdis}. Finally, we conclude the results in \S~\ref{conclu}. In this paper, the photometry is in AB magnitude system 
and we assume $H_{0}=$70 km\,s$^{-1}$\,Mpc$^{-1}$, $\Omega_{m}=$0.3 and $\Omega_{\Lambda}=$0.7.

\section{DATA} \label{data}
\subsection{{\it Cluster Sample}} \label{hecs}

The HeCS clusters were selected from the ROSAT All-Sky Survey \citep{vog99} 
with the Sloan Digital Sky Survey Data Release 6 \citep[SDSS DR6,][]{ade08} footprint. Then, the 
spectroscopic follow-up observation was performed with the Hectospec instrument on MMT. 
Finally, 58 galaxy clusters at $0.1<z<0.3$ were surveyed. From the spectra of mainly 
red-sequence galaxies, \citet{rin13} defined bona fide cluster members through the caustic technique \citep{dia97,dia99} and 
provided cluster information such as redshift, $r_{200}$, $M_{200}$, velocity dispersion ($\sigma_{v}^{\rm cl}$) 
and so on. $r_{200}$ is the radius satisfying that the density is 200 times the critical density and $M_{200}$ 
is a mass within $r_{200}$. The caustic method is advantageous to estimate the cluster mass, because the algorithm 
is based on both galaxy kinematics and positions without assuming dynamical equilibrium, and accurately recovers the mass 
profile up to several Mpc \citep{dia99,ser11}.

To investigate the dependence of member galaxy properties on the central entropy of galaxy clusters, we cross-match 
the HeCS clusters with those in ACCEPT, which 
provides well measured X-ray properties. We mainly adopt the central entropy of the intracluster medium ($K_{0}$) from ACCEPT \citep{cav09}. Based on the X-ray center of 
each cluster, the radial profiles of temperature and electron density were estimated. Then, the radial entropy profile was 
calculated using the estimated temperature and electron density profiles. Finally, the value of $K_{0}$ was 
estimated by fitting the model including a power-law for large radii and a constant value for small radii to the calculated entropy profile. Of 58 HeCS clusters, 
29 clusters overlap with 
the ACCEPT clusters, and we focus on these clusters in this work. Table~\ref{tabinfo} lists 29 clusters and information from HeCS and ACCEPT. \\

\subsection{{\it Spectroscopic/Photometric Data}} \label{photo}

As already mentioned, \citet{rin13} applied the caustic technique to define the members of galaxy clusters. 
Here, we adopt spectroscopic redshift and membership of galaxies 
in galaxy clusters from their result. We complement this data with a spectroscopic sample of galaxies from SDSS DR12 \citep{ala15} and with 
redshifts in the literature \citep[see][for details]{hwa10}. The galaxy clusters 
used in this work have more than 100 members confirmed spectroscopically, except Zw2701 that has 93 members.
Note that the distribution of members spans out to a few Mpc in order to
fully sample the caustic profile.

In addition to the spectroscopic information, photometric information is also important to estimate stellar mass. Since all 
galaxy clusters are covered by the SDSS footprint, we use $ugriz$ model magnitudes to estimate the color of galaxies and 
$r$-band cmodel magnitude for the total magnitude from the SDSS DR12. 
Using the dust map from \citet{sch98}, we also correct the Galactic extinction for each galaxy. Finally, the 
absolute magnitude at $z=$0.1 is calculated with including the $K$-correction \citep{bla07} and 
the evolution correction \citep{teg04} in this work \citep[see also][]{hwa12}.

\subsection{{\it Galaxy Properties}} \label{galprop}

The main purpose of this work is to investigate the association between cluster galaxies and host galaxy clusters. Therefore, 
added to the cluster properties described, it is necessary to use the property of member galaxies such as stellar mass ($M_{*}^{\rm BCG}$) 
and stellar velocity dispersion ($\sigma_{v}^{\rm BCG}$) for BCGs. To gauge the stellar mass of BCGs, we use three different measurements 
for the comparison between different methodologies.

First, to estimate stellar masses of BCGs (see \S~\ref{bcgsel} for BCG identification), we run the Fitting and Assessment of 
Synthetic Templates \citep[FAST,][]{kri09} code with the \citet{con09} stellar population synthesis model. We also assume 
a \citet{cha03} initial mass function, the \citet{cal00} dust attenuation curve, and a delayed star formation history 
(star formation rate $\propto \tau e^{-\tau/t}$). $\rm log\,\tau/yr$ ranges from 8 to 10 with the step size of 0.1 dex, and the age ranges 
from 100 Myr to the universe age of each galaxy with $\Delta \rm log\,t=0.02$. Finally, the internal dust extinction ($A_{V}$) is set between 0 and 5 with 
a 0.05 increment, and the metallicity is allowed to have 0.04, 0.16, 0.51, 1.00 and 1.58 $Z_{\odot}$.

Second, we also use the stellar mass from \citet{mar13} to compare the result with different models. They performed spectral energy 
distribution (SED) fitting with two different templates, passive and star formation ones. Although they use \citet{sal55} and 
\citet{kro01} initial mass functions, we adopt stellar masses derived with a \citet{kro01} 
initial mass function. After cross-matching BCGs with their catalog, we compare $\chi^{2}$ values from both templates, and then 
the stellar masses with smaller $\chi^{2}$ are selected. In total, 23 BCGs are included in their catalog  
with only 6 BCGs (A1068, A1413, A1689, RXJ1504, A2034 and A2259) missing.

Finally, stellar masses and velocity dispersion of BCGs from the MPA/JHU value-added catalog\footnote{http://wwwmpa.mpa-garching.mpg.de/SDSS/DR7/ \\
http://home.strw.leidenuniv.nl/$\sim$jarle/SDSS/} are used as well. The stellar masses were measured by the similar scheme to \citet{kau03}, but 
SDSS photometry with correcting for the contribution of nebular emission was used instead of spectral indices. In this catalog, 21 BCGs are contained. 
Absent BCGs are in A773, A1413, A1423, A1689, A1763, A2034, A2219 and A2259. 
The stellar velocity dispersion in the MPA/JHU catalog is from Princeton/MIT SDSS spectroscopy\footnote{http://spectro.princeton.edu/}. 
Among our BCGs, 19 BCGs are in the catalog with velocity dispersion values measured. The values are listed in Table~\ref{tabbcg}. 
In order to make the values measured by the 
consistent physical scale, we apply the aperture correction with the relation in \citet{mon16}. Also, to minimize the influence of emission lines and 
blue continuum, the effective radius for $z$-band is adopted 
from the New York University Value-added Galaxy Catalog \citep{bla05a,bla05b} for the correction.

\section{RESULTS AND DISCUSSION} \label{resdis}
\subsection{{\it Brightest Members}} \label{bcgcl}

\subsubsection{{\it BCG Identification}} \label{bcgsel}

In order to select BCGs, we initially use the spectroscopic samples only. Among galaxies with spectroscopic redshifts, 
we select the brightest galaxy at $r$-band within a 1 Mpc radius from the X-ray center of each cluster and within the relative rest-frame radial velocity of 
$\pm$ 2000 km\,s$^{-1}$ from the cluster redshift. Then, they are inspected whether there are brighter galaxies from SDSS.
Finally, we define BCGs if there is no brighter galaxy within the velocity range. In the case of A1758, there are two candidates 
with similar brightnesses and redshifts, but one of them is located in the different X-ray peak from that used in ACCEPT. Thus, 
we select the closer one to the X-ray center of ACCEPT. We also note that most BCGs defined are consistent with those in \citet{hof12}, 
except A1758, A1914 and A2069. Their BCGs were identified using near-IR images and redshift information from archives. 
They selected the galaxy in the another X-ray peak for A1758, but we use BCG which is closer to the X-ray center 
in ACCEPT. On the other hand, for the others, our BCGs are brighter than theirs in optical and near-IR bands, and we use the 
latest redshift information from SDSS DR12 and HeCS.

\begin{figure*}
\epsscale{0.6}
\plotone{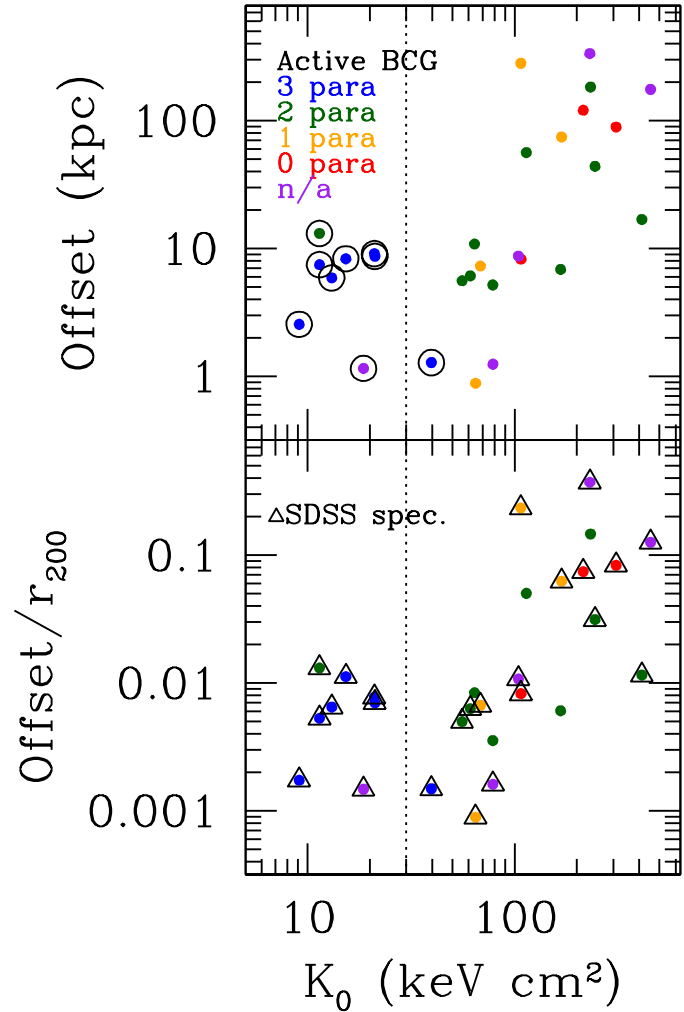}
\caption{The projected X-ray/BCG offset (top) and the offset normalized by cluster radius (bottom) versus central 
entropy for 29 galaxy clusters in this work. Open circles in the top panel indicate BCGs with emission lines. 
The color code presents the number of X-ray morphological parameters satisfying the criteria in \citet{man15}. Open
triangles in the bottom panel are for BCGs with SDSS spectra. 
The dotted line 
shows the value of $K_{0}=$30 keV\,cm$^{2}$ that distinguish galaxy clusters hosting active and passive 
BCGs \citep{cav08}. The BCGs in high $K_{0}$ galaxy clusters tend to be more misaligned. \label{figbcg}}
\end{figure*}

Using the positions of BCGs selected and the X-ray centers from ACCEPT, we calculate the projected offset between those for each cluster. 
The coordinate of BCGs and the calculated offsets are listed in Table~\ref{tabbcg}. 
Figure~\ref{figbcg} displays the X-ray/BCG offset (top) and the offset normalized by the cluster radius (bottom) against the central 
entropy ($K_{0}$) of clusters used in this work. In order to investigate the dependence on the X-ray morphology, we also use the measurement of morphological 
parameters (symmetry, peakiness and alignment) in \citet{man15}. They provided criteria of each parameter to classify galaxy clusters, and the relaxed 
clusters were defined when all parameters satisfy the criteria. Among our sample, 24 clusters have measurements. 
The color code in Figure~\ref{figbcg} represents the number of parameters satisfying their criteria,
i.e., blue points are relaxed clusters in \citet{man15}. The satisfied parameters for each cluster are noted in Table~\ref{tabinfo}. 
The dotted line indicates $K_{0}=$30 keV\,cm$^{2}$ which is known to distinguish galaxy clusters hosting active and passive BCGs \citep{cav08}. 
In fact, from SDSS and literature spectroscopy all 
8 BCGs in galaxy clusters with $K_{0}<$30 keV\,cm$^{2}$ show emission lines, and that in Zw2701 with $K_{0}=$39.66 keV\,cm$^{2}$ also shows the emission lines. 
On the other hand, all 20 BCGs in $K_{0}>$50 keV\,cm$^{2}$ clusters are known to be passive galaxies without emission lines \citep{cav08}.  
Open triangles in the bottom panel denote BCGs 
whose spectra are available from SDSS. Overall, 
it is confirmed that active BCGs with emission lines (open circles in the top panel) are relatively well aligned with the X-ray centers. In addition, all relaxed 
clusters in \citet{man15} have active BCGs. On the other hand, BCGs in galaxy clusters 
with $K_{0}>$30 keV\,cm$^{2}$ show a trend that BCGs in high $K_{0}$ clusters are more misaligned from the X-ray centers. These are in 
good agreement with previous results \citep{kat03,san09,hof12}.

\subsubsection{{\it Dominance of BCGs}} \label{bricl}

The luminosity difference between BCG and the 2nd brightest member is regarded as an indicator of the possibility of recent halo mergers or the formation 
epoch \citep{smi10}. 
Here, we compare the magnitude gap with the central entropy of clusters.

In order to identify 2nd brightest galaxies, we apply the same scheme to the first step for the BCG identification, except using the radius of $r_{200}$ 
and checking galaxies with brightnesses between BCGs and selected 2nd brightest candidates. We also inspect if there are brighter galaxies within the velocity range 
of the caustic profile. From these, we securely identify the 2nd brightest galaxies in 22 galaxy clusters, which are spectroscopic members without other brighter galaxies (group A 
in Table~\ref{tabbcg}). 
In the case of rest of the clusters, there are brighter galaxies than the 2nd brightest ones, but their spectroscopic information is absent. 
For 2nd brightest galaxies without spectroscopic 
redshifts, we use photometric redshift information from SDSS \citep{bec16}. If the redshift difference between the cluster redshift and the photometric redshift is larger 
than 0.129 corresponding to three times the uncertainty of photometric redshifts\footnote{http://www.sdss.org/dr12/algorithms/photo-z/}, we select the original candidate 
from the first step as the 2nd brightest one (group B). On the other hand, if the redshift difference is less than the criterion, the galaxy with the photometric redshift 
is chosen (group C). Then, 
the magnitude gap ($m_{12}$) is defined by $m_{12}=r_{2nd}-r_{\rm BCG}$, where $r_{\rm BCG}$ and $r_{2nd}$ are r-band cmodel magnitudes from SDSS for 
BCG and the 2nd brightest galaxy, respectively. 

\begin{figure*}
\epsscale{0.6}
\plotone{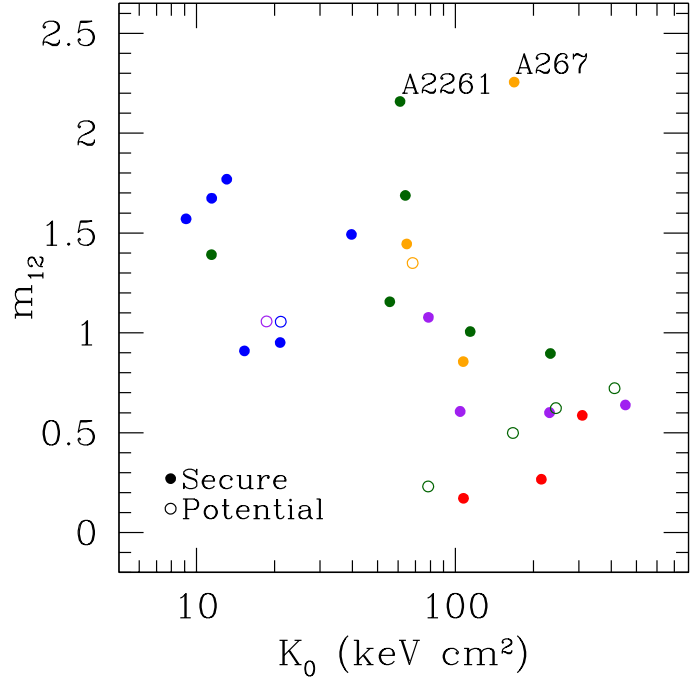}
\caption{The magnitude difference between BCG and the 2nd brightest galaxy against the central entropy of clusters.
The color code is the same to Figure~\ref{figbcg}. 
Filled and open circles indicate securely confirmed 2nd brightest galaxies and potential ones, respectively 
(see the text for more details). The anti-correlation between two parameters appears, i.e., $m_{12}$ gets larger 
as the entropy decreases. We also label two exceptional clusters which do not follow the trend. \label{fig2nd}}
\end{figure*}

Figure~\ref{fig2nd} presents the comparison between the magnitude gap and the cluster central entropy. 
Filled and open symbols indicate secure (group A) and potential (group B and C) 2nd brightest galaxies, respectively. 
The color scheme is the same to Figure~\ref{figbcg}. 
It seems that $m_{12}$ becomes larger when the central entropy gets smaller or clusters are more relaxed, 
which is consistent with published results \citep{smi10,gre16}. Since BCGs in low $K_{0}$ clusters are well aligned with the X-ray center and 
low $K_{0}$ corresponds to a strong cool core, if the central entropy represents the maturity of galaxy clusters, the trends in Figure~\ref{fig2nd} can 
be attributed to the systematic accretion of the most massive cluster members in the cluster core onto the BCG.
In addition, this suggests that $m_{12}$ correlates with the central 
entropy of the intracluster medium of their host galaxy cluster.

In Figure~\ref{fig2nd}, we can also find exceptional clusters which do not follow the trend (A267 and A2261), i.e., the large magnitude gap ($m_{12}>$2) and high entropy. The 
magnitude gap also implies that these clusters are fossil clusters regarded as the most evolved system \citep{jon03}. Assuming that both low entropy and 
the large $m_{12}$ are yardsticks for matured clusters, it is difficult to explain their properties linked to their formation scenarios. 
For example, the halo merger can alters the entropy from low to high. In fact, A267 has 
been classified as merging clusters based on their X-ray morphology, 
although A2261 may be relaxed \citep{zha08}. In \citet{man15}, none of them are classified as relaxed clusters, and 1--2 morphological parameters satisfy 
their criteria. However, each of these two BCGs has a property consistent with a recent massive merger. In A2261, \citet{pos12} 
find an exceptionally large core radius for the BCG with an offset core ``consistent with a local dynamical perturbation of the core''. 
Also, the BCG in A267 shows distinct shells in its outer halo in Hubble Space Telescope imaging\footnote{https://archive.stsci.edu/hst/} 
suggesting a recent galaxy merger. The connection
between $m_{12}$ and the growth of the BCG needs a significantly larger sample to establish a definitive link to mergers,
the central entropy and the dominance of the brightest galaxy. 

\subsection{{\it Red-sequence Members}} \label{rsgal}

Red galaxies in galaxy clusters are mostly composed of old stellar populations. Hence, their colors and magnitudes generate a relation known as 
the red-sequence (RS). However, an age spread of RS galaxies or infalling of new members in the cluster environment can lead to the color scatter on 
the sequence. In this section, we investigate the dependence of RS color scatters on the central entropy of galaxy clusters and the composition of stellar 
population in RS galaxies. For this analysis, BCGs are excluded.

\subsubsection{{\it Color Scatter}} \label{rsscat}

\begin{figure*}
\epsscale{0.6}
\plotone{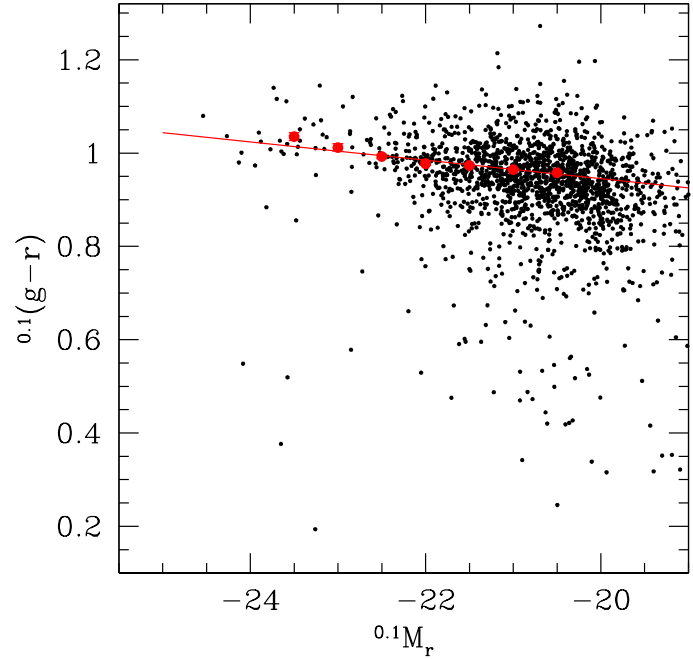}
\caption{The color-magnitude diagram of spectroscopically confirmed member galaxies in 29 galaxy clusters. 
Red points show mean colors in each magnitude bin, and the red line is the defined red-sequence.\label{figrs}}
\end{figure*}

To measure the RS color scatter, we define the red-sequence first. As shown in Figure~\ref{figrs}, we use $K$-corrected colors and absolute magnitude at $z=0.1$ 
of member galaxies in all 29 galaxy clusters. This gives an advantage to avoid the effect of redshift dependence of observed galaxy colors. Mean colors 
of galaxies in each magnitude bin with a 0.5 mag interval are calculated by the 2$\sigma$ clipping algorithm (red points in Figure~\ref{figrs}). Then, we 
perform the linear fit to the calculated colors and central magnitudes of each bin (red line). The best fit RS is $^{0.1}(g-r)=-0.02\,^{0.1}M_{r}+0.55$. 
Finally, the Gaussian distribution is fitted to 
the distribution of color differences between galaxies and the fitted sequence. The final fitting is performed for each galaxy cluster, and we define 
the RS color scatter ($\sigma_{\rm RS}$) with the standard deviation of the Gaussian fit.
We also note that galaxies within 1.5r$_{200}$ from the X-ray center and with $^{0.1}M_{r}<$-20 are used.

\begin{figure*}
\epsscale{0.6}
\plotone{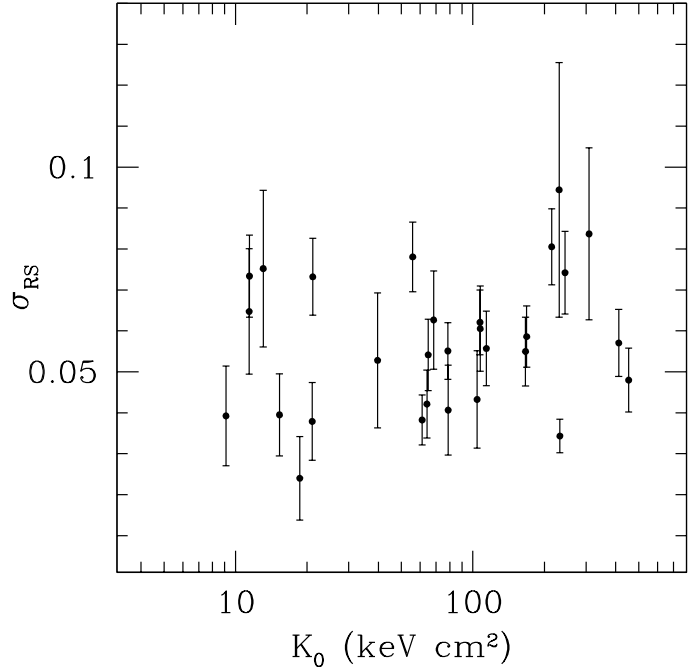}
\caption{The $^{0.1}(g-r)$ scatter of red-sequence galaxies with $M_{r}<-20$ and within 1.5$r_{200}$ from the 
X-ray center. There is no clear dependence of scatters on the central entropy of galaxy clusters.\label{figscat}}
\end{figure*}

Figure~\ref{figscat} shows the scatter against the cluster central entropy. The errors are measured by repeating the bootstrap method 100 times. From 
Figure~\ref{figscat}, there is a cluster-to-cluster variation, but we cannot find significant dependence of $\sigma_{\rm RS}$ on $K_{0}$. The Spearman's 
rank correlation coefficient is 0.22. Furthermore, even if we change the center from the X-ray center to the BCG position, there is still no dependence of 
the color scatter on the central entropy. This implies that the population of RS galaxies is not 
to be affected by the central entropy of the intracluster medium.

\subsubsection{{\it Stellar Population}} \label{rspop}

Previously, we saw that the scatter of RS galaxy colors does not depend on the central entropy of host clusters. Now, we study the stellar population 
of RS galaxies in different central entropy bins.

We use SDSS spectra to obtain the representative spectra of RS galaxies. To select RS galaxies, we apply a criterion of 0.1 mag bluer than the best fit. 
In addition, we use RS galaxies with $^{0.1}M_{r}<-21.9$ in order to 
select RS galaxies with similar masses under the survey depth of SDSS. Using selected RS galaxies, we split them into three groups based on the central 
entropy of host clusters with $K_{0}>$120, 50$<K_{0}<$120 and $K_{0}<$50 keV\,cm$^{2}$. There are 48, 34 and 29 galaxies from the highest to lowest entropy bins. 
Although $K_{0}$ for the last bin is higher than 30 keV\,cm$^{2}$ 
mentioned in the previous section, this bin includes Zw2701, whose BCG also shows emission lines. Therefore, all galaxy clusters in the lowest $K_{0}$ bin 
host active BCGs. Then, spectra of RS galaxies in each bin are stacked after deredshifting to the rest-frame and normalizing with the median flux at $5450\rm \AA<\lambda<5550\AA$. 
The Galactic extinction is also corrected based on $E(B-V)$ from \citet{sch98} and the extinction curve from \citet{car89} with the update for the near-UV 
by \citet{odo94}. 

\begin{figure*}
\epsscale{0.7}
\plotone{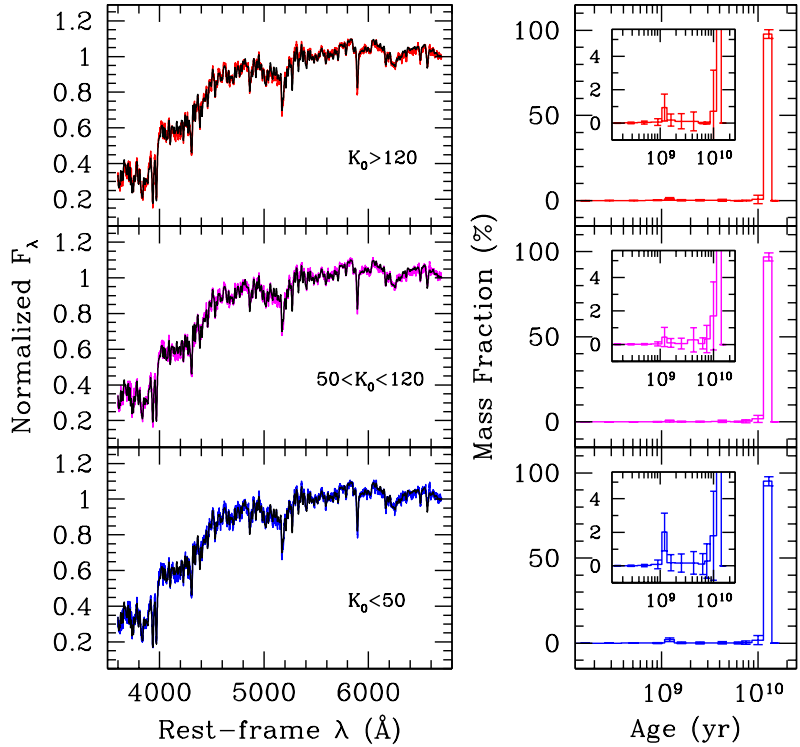}
\caption{The stacked SDSS spectra (left) of red-sequence galaxies with $M_{r}<-21.9$ and within 1.5$r_{200}$ from the 
X-ray center. The galaxy clusters are split into three $K_{0}$ bins, and black lines show examples of best fit results 
from the STARLIGHT code. The right panel shows the stellar mass fraction of stellar populations with different ages. 
The insets are zoomed one showing the fraction of younger populations.
There is no significant difference on the stacked spectrum and stellar population composition.\label{figsp}}
\end{figure*}

Color coded spectra in left panels of Figure~\ref{figsp} show stacked spectra for each bin. To investigate the stellar population, we run STARLIGHT \citep{cid05}. We use the library from 
\citet{bc03} with 6 metallicities (0.005, 0.02, 0.2, 0.4, 1.0, and 2.5 $Z_{\odot}$) and 13 ages (100 Myr -- 13 Gyr). The fitting is repeated 100 times with 100 different random seeds. Black lines in left panels of Figure~\ref{figsp} are best fit examples for each stacked spectrum of RS 
galaxies. The right panels present stellar mass fractions derived by the mean values of 100 fits as a function of age. 
The mass fraction for the oldest bin (age$>$10 Gyr) is over 96 \% for all groups. On the other hand, that for younger stellar population 
(age$<$2.5 Gyr) is 1.3, 0.7 and 2.5 \% for $K_{0}>$120, 50$<K_{0}<$120 and $K_{0}<$50 keV\,cm$^{2}$, respectively. It is clear that 
there is little difference of stellar populations in bright ($^{0.1}M_{r}<-21.9$) RS galaxies with respect to the core properties  of their 
host galaxy cluster, as expected given the relative volume of
the cluster core and the cluster as a whole.  

However, we note that deep near-UV and mid-IR data can help to divide RS members into 
two distinct sub-groups based on the presence of recent star formation. For instance, \citet{ko13,ko16} 
point out that recent star formation traced by near-UV and mid-IR excess is not negligible among
nearby, quiescent early-type galaxies on the tight RS.

\subsection{{\it BCG--Cluster}} \label{bcgcl}

The connection between central galaxies and host halos has been regarded as an important subject, since galaxies form and evolve in their 
host halos. In this section, we investigate the dependence of BCG properties, especially related to their mass, on the properties of 
galaxy clusters. Hereafter, we split BCGs into those in $K_{0}<$50 keV\,cm$^{2}$ clusters (LK-BCGs) and $K_{0}>$50 keV\,cm$^{2}$ clusters (HK-BCGs). 
Noted that all LK-BCGs show emission lines, but none of HK-BCGs show emission lines.

\subsubsection{{\it Stellar Mass to Cluster Mass}} \label{msmh}

\begin{figure*}
\epsscale{0.4}
\plotone{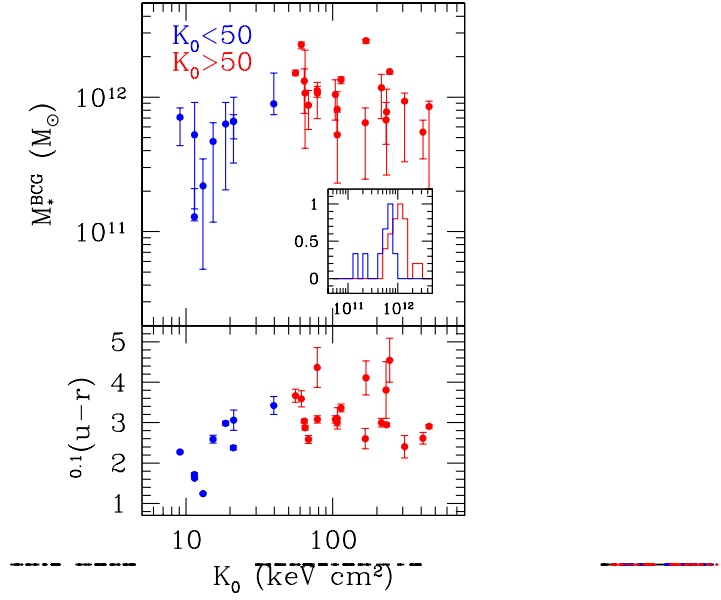}
\caption{Stellar mass of BCGs against the central entropy of galaxy clusters studied in this work (top). Blue and red points are 
BCGs for $K_{0}<$50 and $K_{0}>$50 keV\,cm$^{2}$, respectively. The inset in the top panel shows the stellar mass distribution of each subsample 
normalized with the peak amplitude. The stellar masses of BCGs in low $K_{0}$ clusters are distributed at the low end regime. 
The bottom panel is for $^{0.1}(u-r)$ color of BCGs, which shows BCGs in low $K_{0}$ clusters also tend to be bluer.
\label{figmk}}
\end{figure*}

Figure~\ref{figmk} shows the stellar mass of BCGs ($M_{*}^{\rm BCG}$) we derived against the central entropy of their host clusters (top). The errorbar 
indicates the 1$\sigma$ range of the stellar mass probability distribution. Blue and red points are for LK-BCGs and HK-BCGs, respectively. 
The stellar masses of LK-BCGs are at the low end regime of the stellar mass distribution of our BCGs. The inset in the top panel presents the stellar mass 
distribution of BCGs normalized with the 
peak amplitude of each subsample. Through the Kolmogorov-Smirnov (KS) test, the probability that they are from the same distribution is 0.009. 

\begin{figure*}
\epsscale{0.9}
\plotone{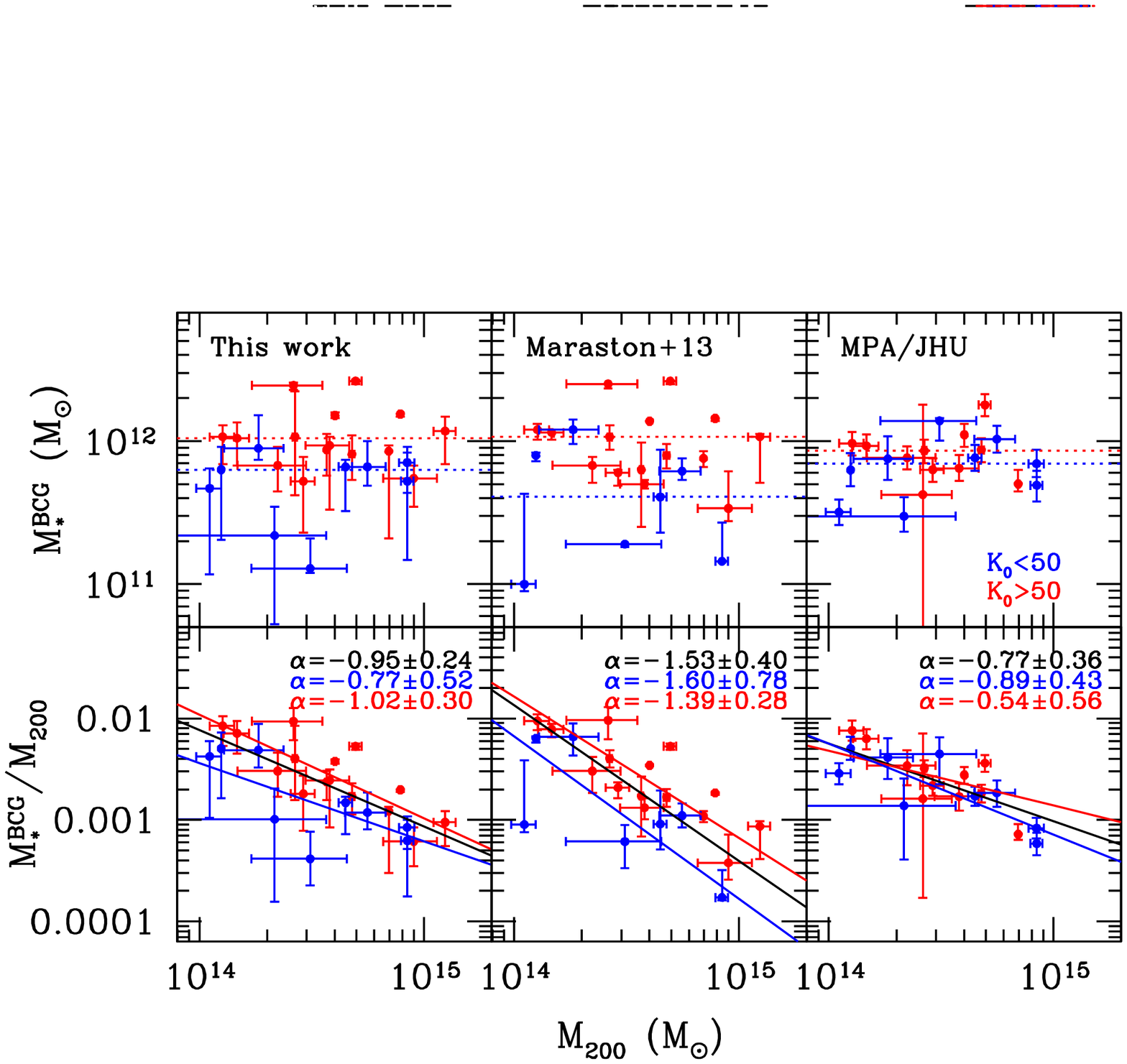}
\caption{Stellar mass (top) and stellar mass-to-cluster mass ratio (bottom) of BCGs against cluster masses. The results of stellar masses 
derived by this work (left), \citet{mar13} (middle) and MPA/JHU (right) are compared. The color scheme is same to Figure~\ref{figmk}. 
Dotted lines in the top panels indicate the median values of each subsample with the same color scheme. 
The stellar mass of BCGs in low entropy clusters can be relatively lower than BCGs in high entropy clusters, although this may be 
affected by how the stellar mass is measured. The solid lines in bottom panels are best fit power-law for all (black), LK-BCGs (blue) 
and HK-BCGs (red). The noted values in each bottom panel are power-law indices of 
$M_{*}^{\rm BCG}/M_{200}\propto M_{200}\,^{\alpha}$ for each subsample.
\label{figmm}}
\end{figure*}

In Figure~\ref{figmm}, we plot $M_{*}^{\rm BCG}$ as a function of cluster mass, $M_{200}$ (top panels). For comparing our measurement (left) with 
other works, we also plot the stellar masses derived by different schemes from \citet{mar13} (middle) and MPA/JHU (right). 
BCGs with spectroscopic information in this work are presented in the left panel, and the cross-matched ones with 
other literatures are shown in middle and right panels. The color scheme is same to Figure~\ref{figmk}. 
Although it seems less significant in the right panel, our result and \citet{mar13} 
commonly show that LK-BCGs (blue) have relatively lower stellar masses than HK-BCGs (red). The offset between median $M_{*}^{\rm BCG}$ values for 
each subsample are 0.22, 0.42 and 0.09 dex for ours, \citet{mar13} and MPA/JHU, respectively. The horizontal dotted lines are median stellar masses.

\citet{gre16} demonstrate that BCGs with strong emission lines can have bluer optical or UV/optical colors and redder mid-IR colors than normal passive BCGs. 
Therefore, this bluer SED by star formation or AGN can lead to lower stellar mass.
In fact, most LK-BCGs in this work also have bluer $u-r$ colors. 
The bottom panel of Figure~\ref{figmk} shows $^{0.1}(u-r)$ color corrected to $z=0.1$ of BCGs as a function of cluster central entropy. It is apparent that 
LK-BCGs are bluer than HK-BCGs. 
In addition, of the eight LK-BCGs included in \citet{kew06}, two are classified as LINERs and the rest 
of them are composite populations from the BPT classification suggesting a mix of star formation and AGN activity.

We also plot the stellar mass to cluster mass ratio ($M_{*}^{\rm BCG}/M_{200}$), instead of $M_{*}^{\rm BCG}$, in the bottom panels of Figure~\ref{figmm}. 
The solid lines are the power-law fit results with the bisector algorithm \citep{iso90} for all (black), LK-BCGs (blue) and HK-BCGs (red). 
The offset is still evident in the ratio between LK-BCGs and HK-BCGs, and the fitted lines also confirm the discrepancy. However, the relation can also depends
on the method used to estimate the stellar mass.
The power-law indices for all BCGs are -0.95$\pm$0.24, -1.53$\pm$0.40 and -0.77$\pm$0.36 from left to right panels. Moreover, from previous studies, it 
was found that the relation between BCG stellar mass and cluster mass is $M_{*}^{\rm BCG} \propto M_{\rm cl}\,^{(0.12\pm0.03)}$ \citep{whi08}, 
$M_{*}^{\rm BCG} \propto M_{500}\,^{(0.78\pm0.06)}$ \citep{sto12} and $M_{*}^{\rm BCG} \propto M_{500}\,^{(0.34\pm0.11)}$ \citep{kra14}, which 
are converted to the power-law index of -0.88, -0.22 and -0.66 for mass ratio and cluster mass. 

\subsubsection{{\it $\sigma^{BCG}$ vs. $\sigma^{cluster}$ Relation}} \label{2sig}

In the previous section, we demonstrated that the stellar mass of BCGs can depend on their current activity and the relation between 
stellar mass and cluster mass can vary depending on the methodology used to derive them. 
In this section, we use another directly observable quantity for an additional comparison, and then discuss what these results imply.

\begin{figure*}
\epsscale{0.6}
\plotone{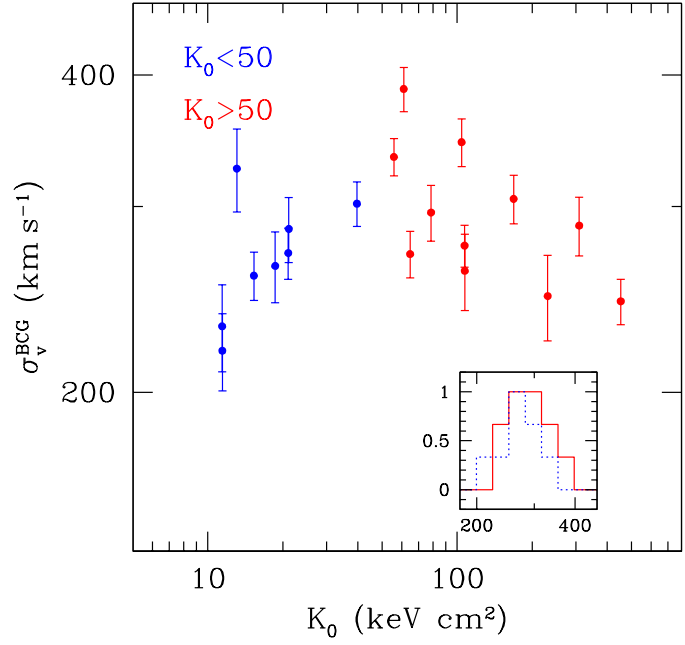}
\caption{Similar figure to Figure~\ref{figmk} with the same color, but the stellar velocity dispersion of BCGs is used. Compared 
to Figure~\ref{figmk}, 
the offset of $\sigma_{v}^{\rm BCG}$ for BCGs in low $K_{0}$ clusters is not significant.
\label{figvk}}
\end{figure*}

Figure~\ref{figvk} is  similar to Figure~\ref{figmk}, but now we use the stellar velocity dispersion of BCGs ($\sigma_{v}^{\rm BCG}$) 
that is the quantity measured directly from the SDSS spectra. Compared to Figure~\ref{figmk}, no dependence to the cluster 
central entropy or the BCG activity is apparent. This is also seen from the inset that shows the similar distribution of 
$\sigma_{v}^{\rm BCG}$ whatever the cluster entropy is. The probability from the KS test is 0.73 which is much higher than 
that for stellar masses in the previous section.

\begin{figure*}
\epsscale{0.8}
\plotone{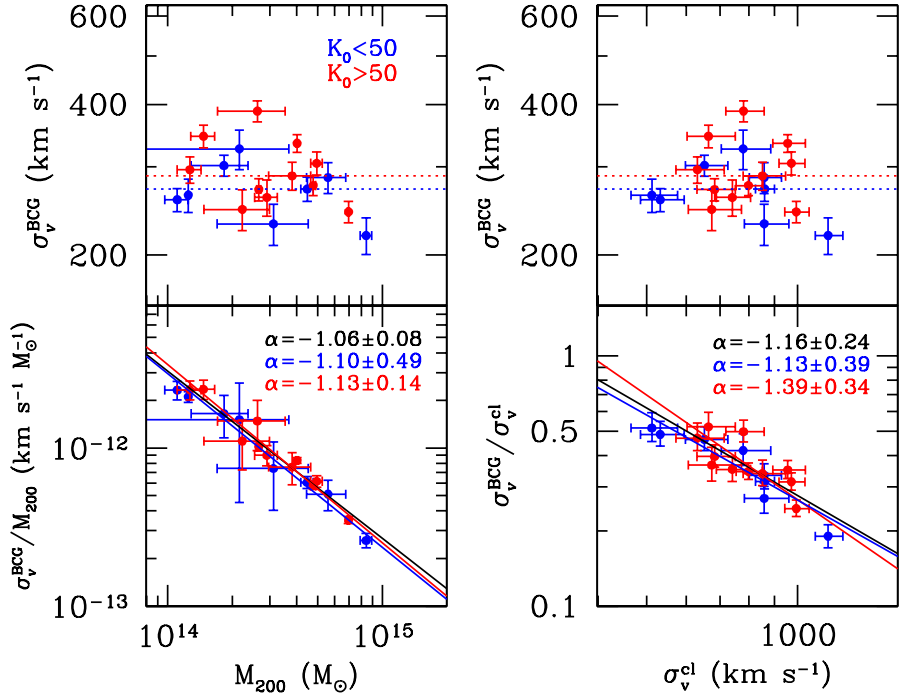}
\caption{Similar figure to Figure~\ref{figmm} with the same color scheme, but the velocity dispersions of BCGs is used 
instead of stellar masses (left). In the right panel, the velocity dispersion of galaxy clusters is used rather than the cluster mass. 
Independently of BCG activity, there is no significant offset between subsamples, especially in terms of the ratio. The noted values in 
each bottom panel are power-law indices of $\sigma_{\rm v}^{\rm BCG}/M_{200}\propto M_{200}\,^{\alpha}$ (left) and 
$\sigma_{\rm v}^{\rm BCG}/\sigma_{\rm v}^{\rm cl}\propto \sigma_{\rm v}^{\rm cl}\,^{\alpha}$ (right) for each subsample.
Noted that BCGs with spectra from SDSS are displayed.
\label{fig2sig}}
\end{figure*}

Again, the top panels of Figure~\ref{fig2sig} show $\sigma_{v}^{\rm BCG}$ against $M_{200}$ (left) and the velocity dispersion of 
host clusters, $\sigma_{v}^{\rm cl}$ (right). It seems that $\sigma_{v}^{\rm BCG}$ of LK-BCGs (blue) is similar to that of 
HK-BCGs (red), which is different from the result based on stellar masses mentioned in the previous section. The difference of 
median $\sigma_{v}^{\rm BCG}$ for subsamples (dotted lines) is 0.03 dex. 
The bottom panels present $\sigma_{v}^{\rm BCG}/M_{200}$--$M_{200}$ (left) and $\sigma_{v}^{\rm BCG}/\sigma_{v}^{\rm cl}$--$\sigma_{v}^{\rm cl}$ (right). 
The power-law indices for all BCGs (black solid line) are -1.06$\pm$0.08 and -1.16$\pm$0.24 for $\sigma_{v}^{\rm BCG}/M_{200}$ and 
$\sigma_{v}^{\rm BCG}/\sigma_{v}^{\rm cl}$, respectively. In addition, the fitted results for LK-BCGs and HK-BCGs are not significantly different. 
Interestingly, the indices for all BCGs are indistinguishable from -1, which may indicate the stellar velocity dispersion of BCGs is nearly constant with a scatter whatever 
cluster masses or velocity dispersions are. However, it is necessary to study the relation with more BCGs over a wider
range of cluster mass for a better constraint.

Differently from the stellar mass, the stellar velocity dispersion is less affected by the dominant light sources such as minor bright 
young stellar populations or AGN. 
In addition, \citet{mcd11} and \citet{mcd16} pointed out that the fuel of star formation in BCGs was galaxy-galaxy interactions at early times, 
but the main source became inctracluster medium cooling recently. Therefore, the similar  
$\sigma^{\rm BCG}_{v}$--$\sigma^{\rm cl}_{v}$ relation irrespective of the BCG activity implies that the bulk of main stellar body or potential 
formed at the early epoch and settled down. However, the current activity imprinting emission lines and color excess 
may be triggered by the intracluster medium recently. Additionally, \citet{ham16} find the rotationally supported gas kinematics at the cluster core 
which is also decoupled from the stellar component of BCGs. In addition, \citet{raw12} reported the external origin of the cold gas 
for BCGs and the star formation fueled by the intracluster medium.

The velocity dispersion  is a directly observable quantity that reflects the gravitational potential of systems. From this work, the velocity dispersion 
seems to be less affected by star formation or AGN activity. 
\citet{wak12a,wak12b} pointed out that the velocity dispersion of galaxies is more tightly 
related to the properties of host dark matter halos and galaxies. Recently, \citet{zah16} also report on the fundamental nature of the central stellar velocity dispersion. 
Furthermore, since the stellar velocity dispersion 
is related to the mass of a supermassive black hole residing in the galaxy center \citep{fer00,geb00}, the relation between $\sigma_{v}^{\rm BCG}$ and 
host cluster also provides an opportunity to link supermassive black holes to the cluster scale dark matter halos as well as 
the relation between central galaxies and their host halos.

Finally, we also note that $\sigma_{v}^{\rm cl}$ and $M_{200}$ for HeCS clusters were mainly derived using red galaxies. However, as \citet{gal08} and \citet{kim16} 
mentioned, there is little evidence that $\sigma_{v}^{\rm cl}$ from red galaxies only or all cluster members are significantly different due to the state
of galaxy clusters at high redshift. 
Therefore, more clusters at various epochs must be studied to understand the evolution of potential wells of BCGs and clusters, and the interrelation between 
central galaxies and their host dark matter halos.

\section{CONCLUSION} \label{conclu}

Using 29 galaxy clusters at 0.1$<z<$0.3 with extensive spectroscopic coverage from HeCS and X-ray information from ACCEPT, we investigated the 
dependence of member galaxy properties on host clusters. The main results are as follows:

\begin{enumerate}

\item Based on BCGs selected and X-ray information, we confirm the connection between the central entropy ($K_{0}$) of clusters and 
X-ray/BCG offset, meaning that BCGs are well aligned in relaxed clusters. Also, the spatial offset between active BCGs and X-ray centers is 
$\lesssim$10 kpc. 

\item The magnitude difference between BCG and the second brightest one ($m_{12}$) is also related to $K_{0}$. BCGs in matured 
clusters are more evolved and become dominant. This also indicates that $m_{12}$ correlates with the central properties of the intracluster medium of the cluster.

\item The color scatter of red-sequence member galaxies does not depend on the central entropy of clusters, but shows a substantial cluster-to-cluster variation. This implies
that the central entropy of the intracluster medium does not influence the red-sequence members, irrespective of the dynamical state of the cluster on larger scales. 

\item BCGs in low entropy clusters (LK-BCGs) showing emission lines are relatively less massive than those in high entropy clusters in terms of the stellar 
component. This  leads to a different 
$M_{*}^{\rm BCG}/M_{200}$--$M_{200}$ relation between BCGs related to their level of activity. The low mass of LK-BCGs may be caused by blue spectral energy 
distribution influenced by minor young massive stars or AGN. Another issue is that different methodologies deriving the stellar mass result in 
different relations dependent on the presence of recent star formation.

\item In contrast to $M_{*}^{\rm BCG}$, the stellar velocity dispersion of BCGs ($\sigma_{v}^{\rm BCG}$) shows no offset between BCGs in high and low entropy clusters. 
This implies that the main stellar body or potential of BCGs have formed earlier, and the activity of LK-BCGs may be recently triggered by other effects such as the 
intracluster medium. 

\end{enumerate}

Here, we used 29 galaxy clusters at relatively low redshift. However, the evolution of galaxy clusters and galaxies is more active in the early universe. 
Hence, it is necessary to use more BCGs at various epochs to understand the evolutionary features. This will provide interesting information 
about the evolution of gravitational potential for galaxies and dark matter halos, and the relation between them. Moreover, since low mass galaxies play a more 
important role to build up the red-sequence \citep[e.g.,][]{del04}, it is also worth to understand the population of low mass cluster galaxies. \\

\acknowledgments

Authors thank anonymous referee for comments improving this manuscript.
A.C.E. acknowledges support from STFC grant ST/L00075X/1.
SDSS-III is managed by the Astrophysical Research Consortium for the
Participating Institutions of the SDSS-III Collaboration including 
the University of Arizona, the Brazilian Participation
Group, Brookhaven National Laboratory, Carnegie Mellon
University, University of Florida, the French Participation
Group, the German Participation Group, Harvard University,
the Instituto de Astrofisica de Canarias, the Michigan State/
Notre Dame/JINA Participation Group, Johns Hopkins University, Lawrence Berkeley National Laboratory, Max Planck 
Institute for Astrophysics, Max Planck Institute for Extraterrestrial Physics, New Mexico State University, New York
University, Ohio State University, Pennsylvania State University, University of Portsmouth, Princeton University, the
Spanish Participation Group, University of Tokyo, University
of Utah, Vanderbilt University, University of Virginia,
University of Washington, and Yale University. 
This research has made use of the NASA/IPAC Extragalactic
Database (NED) which is operated by the Jet Propulsion Laboratory, California
Institute of Technology, under contract with the National Aeronautics and Space
Administration.

\vspace{5mm}
\facilities{SDSS, MMT(Hectospec)}

\software{FAST, STARLIGHT}

\clearpage

\begin{table*}
\begin{center}
\caption{Summary of galaxy clusters used in this work.
\label{tabinfo}}
\begin{tabular}{ccccccccc}
\tableline\tableline
Cluster & \multicolumn{2}{c}{X-ray Coordinates$^{a}$} &  $z$$^{b}$ & $r_{200}$$^{b}$ & $M_{200}$$^{b}$ & $\sigma_{v}^{\rm cl}$\,$^{b}$ & $K_{0}$$^{a}$ & Morph.$^{c}$\\
        & R.A. (J2000)               & decl. (J2000)  &     &      (Mpc)      & (10$^{14} M_{\odot}$) & (km\,s$^{-1}$) & (keV cm$^{2}$) & \\
\tableline
A267       & 01:52:42.27  & +01:00:45.33 &  0.2291 & 1.19 & 4.95$\pm$0.31   & 972$^{+63}_{-53}$  & 168.56 & A \\  
A697       & 08:42:57.55  & +36:21:57.65 &  0.2812 & 1.13 & 4.42$\pm$2.10   & 1002$^{+97}_{-75}$ & 166.67 & SA \\ 
MS0906     & 09:09:12.75  & +10:58:32.00 &  0.1767 & 0.81 & 1.47$\pm$0.19   & 664$^{+87}_{-62}$  & 104.23 & n/a \\
A773       & 09:17:52.57  & +51:43:38.18 &  0.2173 & 1.40 & 7.84$\pm$0.10   & 1110$^{+86}_{-70}$ & 244.32 & SA \\ 
Zw2701     & 09:52:49.18  & +51:53:05.27 &  0.2160 & 0.86 & 1.83$\pm$0.54   & 652$^{+74}_{-55}$  &  39.66 & SPA \\
A963       & 10:17:03.74  & +39:02:49.17 &  0.2041 & 1.12 & 4.01$\pm$0.05   & 956$^{+80}_{-64}$  &  55.77 & SA \\ 
Zw3146     & 10:23:39.74  & +04:11:08.05 &  0.2894 & 1.00 & 3.11$\pm$1.41   & 858$^{+103}_{-75}$ & 11.42  & PA \\ 
A1068      & 10:40:44.52  & +39:57:10.28 &  0.1386 & 1.47 & 8.40$\pm$0.66   & 1028$^{+106}_{-81}$ & 9.11  & SPA \\
A1201      & 11:12:54.49  & +13:26:08.76 &  0.1671 & 0.99 & 2.66$\pm$0.06   & 683$^{+68}_{-53}$ &  64.81  & A \\  
A1204      & 11:13:20.42  & +17:35:38.45 &  0.1706 & 0.74 & 1.11$\pm$0.14   & 532$^{+62}_{-46}$ &  15.31  & SPA \\
A1361      & 11:43:39.64  & +46:21:20.41 &  0.1159 & 0.78 & 1.25$\pm$0.00   & 512$^{+64}_{-47}$ &  18.64  & n/a \\
A1413      & 11:55:17.89  & +23:24:21.84 &  0.1412 & 1.29 & 5.72$\pm$0.02   & 856$^{+90}_{-68}$ &  64.03  & SA \\ 
A1423      & 11:57:17.26  & +33:36:37.44 &  0.2142 & 1.09 & 3.68$\pm$0.06   & 759$^{+64}_{-51}$ &  68.32  & S \\  
A1689      & 13:11:29.61  & -01:20:28.69 &  0.1842 & 1.46 & 8.68$\pm$2.64   & 1197$^{+78}_{-65}$ &  78.44 & SA \\ 
A1758      & 13:32:48.40  & +50:32:32.53 &  0.2760 & 0.90 & 2.23$\pm$0.75   & 674$^{+99}_{-69}$ & 230.84  & n/a \\
A1763      & 13:35:17.96  & +40:59:55.80 &  0.2312 & 1.62 & 12.40$\pm$1.39  & 1261$^{+81}_{-68}$ & 214.69 & 0 \\  
A1835      & 14:01:01.95  & +02:52:43.18 &  0.2506 & 1.41 & 8.41$\pm$0.53   & 1151$^{+80}_{-66}$ &  11.44 & SPA \\
A1914      & 14:26:03.06  & +37:49:27.84 &  0.1660 & 1.20 & 4.77$\pm$0.13   & 798$^{+53}_{-44}$ & 107.16  & A \\  
RXJ1504    & 15:04:07.42  & -02:48:15.70 &  0.2168 & 0.91 & 2.16$\pm$1.51   & 779$^{+105}_{-75}$ &  13.08 & SPA \\
A2034      & 15:10:12.50  & +33:30:39.57 &  0.1132 & 1.25 & 5.03$\pm$0.05   & 942$^{+64}_{-53}$ & 232.64  & SA \\ 
A2069      & 15:24:11.38  & +29:52:19.02 &  0.1139 & 1.39 & 6.96$\pm$0.08   & 994$^{+61}_{-52}$ & 453.25  & n/a \\
A2111      & 15:39:40.64  & +34:25:28.01 &  0.2291 & 1.00 & 2.90$\pm$0.35   & 741$^{+65}_{-52}$ & 107.36  & 0 \\  
A2187      & 16:24:14.02  & +41:14:37.53 &  0.1829 & 0.77 & 1.27$\pm$0.16   & 631$^{+83}_{-59}$ &  78.63  & n/a \\
A2219      & 16:40:20.11  & +46:42:42.84 &  0.2257 & 1.46 & 8.98$\pm$2.42   & 1151$^{+63}_{-54}$ & 411.57 & SA \\ 
A2259      & 17:20:08.30  & +27:40:11.53 &  0.1605 & 1.12 & 3.84$\pm$0.68   & 855$^{+76}_{-60}$ & 113.98  & SA \\ 
RXJ1720    & 17:20:09.94  & +26:37:29.11 &  0.1604 & 1.18 & 4.47$\pm$0.30   & 860$^{+40}_{-35}$ &  21.03  & SPA \\
A2261      & 17:22:27.25  & +32:07:58.60 &  0.2242 & 0.97 & 2.62$\pm$0.91   & 780$^{+78}_{-60}$ &  61.08  & SA \\ 
RXJ2129    & 21:29:39.94  & +00:05:18.83 &  0.2339 & 1.24 & 5.59$\pm$1.16   & 858$^{+71}_{-57}$ &  21.14  & SPA \\
A2631      & 23:37:38.56  & +00:16:05.02 &  0.2765 & 1.07 & 3.80$\pm$0.84   & 851$^{+96}_{-72}$ & 308.81  & 0 \\  
\tableline
\end{tabular}
\end{center}
\tablenotetext{a}{Values based on the ACCEPT data \citep{cav09}}
\tablenotetext{b}{Values based on the HeCS data \citep{rin13}}
\tablenotetext{c}{X-ray morphological parameters satisfying the criteria in \citet{man15}. The parameters of S, P and A 
indicate symmetry, peakiness and alignment, respectively. 0 means that all parameters do not satisfy the criteria. 
If there is no measurement from \citet{man15}, we note n/a.}
\end{table*}


\begin{table*}
\begin{center}
\caption{Summary of BCGs and the 2nd brightest galaxies defined in this work. Column (2-3) give the coordinate of BCGs, and column (4) lists the offset between positions of BCG and 
the X-ray center. Column (5) is the stellar mass derived in this work. Column (6) is the stellar velocity dispersion of BCGs described in 
\S~\ref{galprop}. Column (7-8) give the coordinate of the 2nd brightest galaxies, and column (9) is the magnitude difference between 
BCGs and the 2nd brightest galaxies. The final column indicates how the 2nd brightest one was selected (see the text). It is noted that we do not 
present values if they are unavailable. 
\label{tabbcg}}
\begin{tabular}{cccccccccc}
\tableline\tableline
Cluster & \multicolumn{2}{c}{BCG Coordinates} &  Offset & log$M_{*}^{\rm BCG}$ & $\sigma_{v}^{\rm BCG}$ & \multicolumn{2}{c}{2nd Brightest} & $m_{12}$ & Group \\
        & R.A. (J2000)       & decl. (J2000)  &   (kpc) & ($M_{\odot}$) &    (km\,s$^{-1}$)       &  R.A. (J2000) & decl. (J2000) & (mag) & \\
\tableline
A267       & 01:52:42  & +01:00:26 &  74.46   & 12.42$^{+0.01}_{-0.02}$ & 297$\pm$16 & 01:52:22 & +01:00:08 & 2.2558 & A \\
A697       & 08:42:58  & +36:21:59 &   6.85   & 11.81$^{+0.11}_{-0.42}$ & -          & 08:42:58 & +36:22:01 & 0.4992 & B \\
MS0906     & 09:09:13  & +10:58:29 &   8.71   & 12.02$^{+0.11}_{-0.19}$ & 341$\pm$18 & 09:09:07 & +10:57:51 & 0.6064 & A \\
A773       & 09:17:53  & +51:43:37 &  43.91   & 12.19$^{+0.00}_{-0.02}$ & -          & 09:17:53 & +51:44:01 & 0.6228 & B \\
Zw2701     & 09:52:49  & +51:53:05 &   1.28   & 11.95$^{+0.23}_{-0.08}$ & 298$\pm$15 & 09:53:01 & +51:52:25 & 1.4925 & A \\
A963       & 10:17:04  & +39:02:49 &   5.58   & 12.18$^{+0.02}_{-0.02}$ & 330$\pm$14 & 10:17:22 & +39:00:07 & 1.1557 & A \\
Zw3146     & 10:23:40  & +04:11:11 &  13.12   & 11.11$^{+0.21}_{-0.03}$ & 229$\pm$22 & 10:23:37 & +04:09:06 & 1.3916 & A \\
A1068      & 10:40:44  & +39:57:11 &   2.55   & 11.85$^{+0.07}_{-0.21}$ & -          & 10:40:34 & +40:03:49 & 1.5706 & A \\
A1201      & 11:12:55  & +13:26:09 &   0.88   & 12.03$^{+0.32}_{-0.41}$ & 267$\pm$14 & 11:12:50 & +13:28:30 & 1.4455 & A \\
A1204      & 11:13:21  & +17:35:41 &   8.29   & 11.67$^{+0.14}_{-0.60}$ & 260$\pm$14 & 11:13:32 & +17:38:42 & 0.9101 & A \\
A1361      & 11:43:40  & +46:21:20 &   1.15   & 11.80$^{+0.16}_{-0.49}$ & 262$\pm$20 & 11:44:00 & +46:24:23 & 1.0571 & C \\
A1413      & 11:55:18  & +23:24:18 &  10.83   & 12.12$^{+0.09}_{-0.24}$ & -          & 11:54:58 & +23:25:20 & 1.6880 & A \\
A1423      & 11:57:17  & +33:36:39 &   7.29   & 11.94$^{+0.11}_{-0.18}$ & -          & 11:57:36 & +33:34:39 & 1.3500 & C \\
A1689      & 13:11:30  & -01:20:28 &   5.18   & 12.05$^{+0.03}_{-0.05}$ & -          & 13:11:30 & -01:20:43 & 0.2307 & C \\
A1758      & 13:32:52  & +50:31:34 & 334.75   & 11.83$^{+0.13}_{-0.18}$ & 245$\pm$23 & 13:32:41 & +50:33:46 & 0.6007 & A \\
A1763      & 13:35:20  & +41:00:04 & 120.44   & 12.07$^{+0.10}_{-0.23}$ & -          & 13:34:54 & +40:56:55 & 0.2665 & A \\
A1835      & 14:01:02  & +02:52:42 &   7.46   & 11.72$^{+0.24}_{-0.55}$ & 221$\pm$18 & 14:01:07 & +02:50:55 & 1.6744 & A \\
A1914      & 14:25:57  & +37:48:59 & 280.82   & 11.91$^{+0.13}_{-0.18}$ & 274$\pm$13 & 14:26:04 & +37:49:53 & 0.8563 & A \\
RXJ1504    & 15:04:08  & -02:48:17 &   5.89   & 11.34$^{+0.20}_{-0.62}$ & 326$\pm$30 & 15:04:23 & -02:47:29 & 1.7690 & A \\
A2034      & 15:10:12  & +33:29:11 & 183.11   & 11.89$^{+0.17}_{-0.47}$ & -          & 15:10:20 & +33:29:10 & 0.8965 & A \\
A2069      & 15:24:07  & +29:53:20 & 175.46   & 11.93$^{+0.04}_{-0.61}$ & 241$\pm$12 & 15:24:08 & +29:52:55 & 0.6387 & A \\
A2111      & 15:39:40  & +34:25:27 &   8.25   & 11.72$^{+0.17}_{-0.36}$ & 261$\pm$22 & 15:39:42 & +34:24:43 & 0.1720 & A \\
A2187      & 16:24:14  & +41:14:38 &   1.24   & 12.03$^{+0.08}_{-0.19}$ & 292$\pm$18 & 16:24:23 & +41:15:37 & 1.0774 & A \\
A2219      & 16:40:20  & +46:42:41 &  16.86   & 11.74$^{+0.09}_{-0.20}$ & -          & 16:40:32 & +46:42:30 & 0.7224 & C \\
A2259      & 17:20:10  & +27:40:08 &  56.25   & 12.13$^{+0.02}_{-0.03}$ & -          & 17:20:11 & +27:42:14 & 1.0062 & A \\
RXJ1720    & 17:20:10  & +26:37:32 &   9.12   & 11.82$^{+0.05}_{-0.31}$ & 273$\pm$15 & 17:20:32 & +26:40:20 & 0.9514 & A \\
A2261      & 17:22:27  & +32:07:57 &   6.11   & 12.39$^{+0.02}_{-0.03}$ & 386$\pm$19 & 17:22:35 & +32:07:44 & 2.1591 & A \\
RXJ2129    & 21:29:40  & +00:05:21 &   8.68   & 11.82$^{+0.18}_{-0.13}$ & 285$\pm$20 & 21:29:36 & +00:01:29 & 1.0558 & C \\
A2631      & 23:37:40  & +00:16:17 &  89.11   & 11.97$^{+0.06}_{-0.45}$ & 289$\pm$18 & 23:37:24 & +00:16:21 & 0.5863 & A \\
\tableline
\end{tabular}
\end{center}
\end{table*}

\clearpage

\begin{thebibliography}{}

\bibitem[Adelman-McCarthy et al. (2008)]{ade08} Adelman-McCarthy, J. K., Ag\"{u}eros, M. A., Allam, S. S., et al. 2008, \apjs, 175, 297
\bibitem[Alam et al. (2015)]{ala15} Alam, S., Albareti, F. D., Allende Prieto, C., et al. 2015, \apjs, 219, 12
\bibitem[Baugh (2006)]{bau06} Baugh, C. M. 2006, RPPh, 69, 3101
\bibitem[Beck et al. (2016)]{bec16} Beck, R., Dobos, L., Budav\'{a}ri, T., Szalay, A. S., \& Csabai, I. 2016, \mnras, 460, 1371
\bibitem[Behroozi et al. (2010)]{beh10} Behroozi, P. S., Conroy, C., \& Wechsler, R. H. 2010, \apj, 717, 379
\bibitem[Blanton et al. (2005a)]{bla05a} Blanton, M. R., Eisenstein, D., Hogg, D. W., Schlegel, D. J., \& Brinkmann, J. 2005a, \apj, 629, 143
\bibitem[Blanton \& Roweis (2007)]{bla07} Blanton, M. R., \& Roweis, S. 2007, \aj, 133, 734
\bibitem[Blanton et al. (2005b)]{bla05b} Blanton, M. R., Schlegel, D. J., Strauss, M. A., et al. 2005b, \aj, 129, 2562
\bibitem[Bruzual \& Charlot (2003)]{bc03} Bruzual, G., \& Charlot, S. 2003, \mnras, 344, 1000
\bibitem[Calzetti et al. (2000)]{cal00} Calzetti, D., Armus, L., Bolhlin, R. C., et al. 2000, \apj, 533, 682
\bibitem[Cardelli et al. (1989)]{car89} Cardelli, J. A., Clayton, G. C., \& Mathis, J. S. 1989, \apj, 345, 245
\bibitem[Cavagnolo et al. (2008)]{cav08} Cavagnolo, K. W., Donahue, M., Voit, G. M., \& Sun, M. 2008, \apjl, 683, L107
\bibitem[Cavagnolo et al. (2009)]{cav09} Cavagnolo, K. W., Donahue, M., Voit, G. M., \& Sun, M. 2009, \apj, 182, 12
\bibitem[Chabrier (2003)]{cha03} Chabrier, G. 2003, \pasp, 115, 763
\bibitem[Cid Fernandes et al. (2005)]{cid05} Cid Fernandes, R., Mateus, A., Sodr\'{e}, L., Stasi\'{n}ska, G., \& Gomes, J. M. 2005, \mnras, 358, 363
\bibitem[Conroy et al. (2009)]{con09} Conroy, C., Gunn, J. E., \& White, M. 2009, \apj, 699, 486
\bibitem[Crawford et al. (1999)]{cra99} Crawford, C. S., Allen, S. W., Ebeling, H., Edge, A. C., \& Fabian, A. C. 1999, \mnras, 306, 857
\bibitem[De Lucia et al. (2004)]{del04} De Lucia, G., Poggianti, B. M., Arag\'{o}n-Salamanca, A., et al. 2004, \apjl, 610, L77
\bibitem[Donahue et al. (2010)]{don10} Donahue, M., Bruch, S., Wang, E., et al. 2010, \apj, 715, 881
\bibitem[Diaferio (1999)]{dia99} Diaferio, A. 1999, \mnras, 309, 610
\bibitem[Diaferio \& Geller (1997)]{dia97} Diaferio, A., \& Geller, M. J. 1997, \apj, 481, 633
\bibitem[Edge (2001)]{edg01} Edge, A. C. 2001, \mnras, 328, 762
\bibitem[Edge et al. (1999)]{edg99} Edge, A. C., Ivison, R. J., Smail, I., Blain, A. W., \& Kneib, J. -P. 1999, \mnras, 306, 599
\bibitem[Edge et al. (2010)]{edg10} Edge, A. C., Oonk, J. B. R., Mittal, R., et al. 2010, \aap, 518, L46
\bibitem[Edge et al. (2002)]{edg02} Edge, A. C., Wilman, R. J., Johnstone, R. M., et al. 2002, \mnras, 337, 49
\bibitem[Egami et al. (2006)]{ega06} Egami, E., Rieke, G. H., Fadda, D., \& Hines, D. C. 2006, \apjl, 652, L21
\bibitem[Ferrarese \& Merritt (2000)]{fer00} Ferrarese, L., \& Merritt, D. 2000, \apjl, 539, L9
\bibitem[Fogarty et al. (2015)]{fog15} Fogarty, K., Postman, M., Connor, T., Donahue, M., \& Moustakas, J. 2015, \apj, 813, 117
\bibitem[Gal et al. (2008)]{gal08} Gal, R. R., Lemaux, B. C., Lubin, L. M., Kocevski, D., \& Squires, G. K. 2008, \apj, 684, 933
\bibitem[Gebhardt et al. (2000)]{geb00} Gebhardt, K., Bender, R., Bower, G., et al. 2000, \apjl, 539, L13
\bibitem[Gonzalez et al. (2007)]{gon07} Gonzalez, A. H., Zaritsky, D., \& Zabludoff, A. I. 2007, \apj, 666, 147
\bibitem[Green et al. (2016)]{gre16} Green, T. S., Edge, A. C., Stott, J. P., et al. 2016, \mnras, 461, 560
\bibitem[Groenewald \& Loubser (2014)]{gro14} Groenewald, D. N., \& Loubser, S. I. 2014, \mnras, 444, 808
\bibitem[Hamer et al. (2016)]{ham16} Hamer, S. L., Edge, A. C., Swinbank, A. M., et al. 2016, \mnras, 460, 1758
\bibitem[Hansen et al. (2009)]{han09} Hansen, S. M., Sheldon, E. S., Wechsler, R. H., \& Koester, B. P. 2009, \apj, 699, 1333
\bibitem[Hashimoto et al. (2014)]{has14} Hashimoto, Y., Henry, J. P., \& Boehringer, H. 2014, \mnras, 440, 588
\bibitem[Hoffer et al. (2012)]{hof12} Hoffer, A. S., Donahue, M., Hicks, A., \& Barthelemy, R. S. 2012, \apjs, 199, 23
\bibitem[Hwang et al. (2016)]{hwa16} Hwang, H. S., Geller, M. J., Park, C., et al. 2016, \apj, 818, 173
\bibitem[Hwang et al. (2010)]{hwa10} Hwang, H. S., Elbaz, D., Lee, J. C., Jeong, W.-S., Park, C., Lee, M. G., \& Lee, H. M. 
2010, \aap, 522, 33
\bibitem[Hwang et al. (2012)]{hwa12} Hwang, H. S., Park, C., Elbaz, D., \& Choi, Y. -Y. 2012, \aap, 538, 15
\bibitem[Isobe et al. (1990)]{iso90} Isobe, T., Feigelson, E. D., Akritas, M. G., \& Babu, G. J. 1990, \apj, 364, 104
\bibitem[Jones et al. (2003)]{jon03} Jones, L. R., Ponman, T. J., Horton, A., Babul, A., Ebeling, H., \& Burke, D. J. 2003, 
\mnras, 343, 627
\bibitem[Katayama et al. (2003)]{kat03} Katayama, H., Hayashida, K., Takahara, F., \& Fujita, Y. 2003, \apj, 585, 687
\bibitem[Kauffmann et al. (2003)]{kau03} Kauffmann, G., Heckman, T. M., White, S. D. M., et al. 2003, \mnras, 341, 33
\bibitem[Kewley et al. (2006)]{kew06} Kewley, L. J., Groves, B., Kauffmann, G., \& Heckman, T. 2006, \mnras, 372, 961
\bibitem[Kravtsov et al. (2014)]{kra14} Kravtsov, A., Vikhlinin, A., \& Meshscheryakov, A. 2014, arXiv:1401.7329
\bibitem[Kim et al. (2015)]{kim15} Kim, J. -W., Im, M., Lee, S. -K., et al. 2015, \apj, 806, 189
\bibitem[Kim et al. (2016)]{kim16} Kim, J. -W., Im, M., Lee, S. -K., et al. 2016, \apjl, 821, L10
\bibitem[Kriek et al. (2009)]{kri09} Kriek, M., van Dokkum, P. G., Labb\'{e}, I., et al. 2009, \apj, 700, 221
\bibitem[Ko et al. (2013)]{ko13} Ko, J., Hwang, H. S., Lee, J. C., \& Sohn, Y. -J. 2013, \apj, 767, 90
\bibitem[Ko et al. (2016)]{ko16} Ko, J., Chung, H., Hwang, H. S., \& Lee, J. C. 2016, \apj, 820, 2
\bibitem[Kroupa (2001)]{kro01} Kroupa, P. 2001, \mnras, 322, 231
\bibitem[Lauer et al. (2014)]{lau14} Lauer, T. R., Postman, M., Strauss, M. A., Graves, G. J., \& Chisari, N. E. 2014, \apj, 797, 82
\bibitem[Lin \& Mohr (2004)]{lin04} Lin, Y. -T., \& Mohr, J. J. 2004, \apj, 617, 879
\bibitem[Loh \& Strauss (2006)]{loh06} Loh, Y. -S., \& Strauss, M. A. 2006, \mnras, 366, 373
\bibitem[Mantz et al. (2015)]{man15} Mantz, A. B., Allen, S. W., Morris, R. G., Schmidt, R. W., von der Linden, A., \& 
Urban, O. 2015, \mnras, 449, 199
\bibitem[Maraston et al. (2013)]{mar13} Maraston, C., Pforr, J., Henriques, B. M., et al. 2013, \mnras, 435, 2764
\bibitem[McDonald et al. (2016)]{mcd16} McDonald, M., Stalder, B., Bayliss, M., et al. 2016, \apj, 817, 86
\bibitem[McDonald (2011)]{mcd11} McDonald, M. 2011, \apjl, 742, L35
\bibitem[Mittal et al. (2012)]{mit12} Mittal, R., Oonk, J. B. R., Ferland, G. J., et al. 2012, \mnras, 426, 2957
\bibitem[Montero-Dorta et al. (2016)]{mon16} Montero-Dorta, A. D., Shu, Y., Bolton, A. S., Brownstein, J. R., \& Weiner, B. J. 
2016, \mnras, 456, 3265
\bibitem[Moster et al. (2010)]{mos10} Moster, B. P., Somerville, R. S., Maulbetsch, C., et al. 2010, \apj, 710, 903
\bibitem[Pipino et al. (2011)]{pip11} Pipino, A., Szabo, T., Pierpaoli, E., MacKenzie, S. M., \& Dong, F. 2011, \mnras, 417, 2817
\bibitem[Postman et al. (2012)]{pos12} Postman, M., Lauer, T. R., Donahue, M., et al. 2012, \apj, 756, 159
\bibitem[O'Donnell (1994)]{odo94} O'Donnell, J. E. 1994, \apj, 422, 158
\bibitem[Ostriker \& Tremaine (1975)]{ost75} Ostriker, J. P., \& Tremaine, S. D. 1975, \apjl, 202, L113
\bibitem[Rafferty et al. (2008)]{raf08} Rafferty, D. A., McNamara, B. R., \& Nulsen, P. E. J. 2008, \apj, 687, 899
\bibitem[Rawle et al. (2012)]{raw12} Rawle, T. D., Edge, A. C., Egami, E., et al., 2012, \apj, 747, 29
\bibitem[Rines et al. (2013)]{rin13} Rines, K., Geller, M. J., Diaferio, A., \& Kurtz, M. J. 2013, \apj, 767, 15
\bibitem[Salom\'{e} \& Combes (2003)]{sal03} Salom\'{e}, P., \& Combes, F. 2003, \aap, 412, 657
\bibitem[Salpeter (1955)]{sal55} Salpeter, E. E. 1955, \apj, 121, 161
\bibitem[Sanderson et al. (2009)]{san09} Sanderson, A. J. R., Edge, A. C., \& Smith, G. P. 2009, \mnras, 398, 1698
\bibitem[Schlegel et al. (1998)]{sch98} Schlegel, D. J., Finkbeiner, D. P., \& David, M. 1998, \apj, 500, 525
\bibitem[Serra et al. (2011)]{ser11} Serra, A. L., Diaferio, A., Murante, G., \& Borgani, S. 2011, \mnras, 412, 800
\bibitem[Smith et al. (2010)]{smi10} Smith, G. P., Khosroshahi, H. G., Dariush, A., et al. 2010, \mnras, 409, 169
\bibitem[Stott et al. (2012)]{sto12} Stott, J. P., Hickox, R. C., Edge, A. C., et al. 2012, \mnras, 422, 2213
\bibitem[Tegmark et al. (2004)]{teg04} Tegmark, M., Blanton, M. R., Strauss, M. A., et al. 2004, \apj, 606, 702
\bibitem[Voges et al. (1999)]{vog99} Voges, W., Aschenbach, B., Boller, Th., et al. 1999, \aap, 349, 389
\bibitem[Wake et al. (2012a)]{wak12a} Wake, D. A., Franx, M., \& van Dokkum, P. G. 2012a, arXiv:1201.1913
\bibitem[Wake et al. (2012b)]{wak12b} Wake, D. A., van Dokkum, P. G., \& Franx, M. 2012b, \apjl, 751, L44
\bibitem[Wake et al. (2011)]{wak11} Wake, D. A., Whitaker, K. E., Labb\'{e}, I., et al. 2011, \apj, 728, 46
\bibitem[Wang et al. (2010)]{wan10} Wang, J., Overzier, R., Kauffmann, G., von der Linde, A., \& Kong, X. 2010, \mnras, 401, 433
\bibitem[Whiley et al. (2008)]{whi08} Whiley, I. M., Arag\'{o}n-Salamanca, A., De Lucia, G., et al. 2008, \mnras, 387, 1253
\bibitem[White \& Rees (1978)]{whi78} White, S. D. M., \& Rees, M. J. 1978, \mnras, 183, 341
\bibitem[Zahid et al. (2016)]{zah16} Zahid, H. J., Geller, M. J., Fabricant, D. G., \& Hwang, H. S. 2016, \apj, 832, 203
\bibitem[Zhang et al. (2008)]{zha08} Zhang, Y. Y., Finoguenov, A., B\"{o}hringer, H., Kneib, J. -P., Smith, G. P., Kneissl, R., 
Okabe, N., \& Dahle, H. 2008, \aap, 482, 451
\bibitem[Zheng et al. (2007)]{zhe07} Zheng, Z., Coil, A. L., \& Zehavi, I. 2007, \apj, 667, 760

\end{thebibliography}
\end{document}